%% file: ms.tex
\input{psfig.sty}
\documentclass[12pt,preprint]{aastex}
\usepackage{graphicx}

\shorttitle{Optical Depth in stellar coronae}
\shortauthors{Testa et al.}
\usepackage{amsmath}

\begin{document}
\title{Detection of X-ray Resonance Scattering in Active Stellar Coronae}
\author{Paola Testa\altaffilmark{1,2,3}, Jeremy
J. Drake\altaffilmark{2}, Giovanni Peres\altaffilmark{3} and Edward
E.\ DeLuca\altaffilmark{2}}
\altaffiltext{1}{SAO Predoctoral Fellow}
\altaffiltext{2}{Smithsonian Astrophysical Observatory, MS 3, 60 Garden
Street, Cambridge, MA 02138, USA; jdrake@head.cfa.harvard.edu}
\altaffiltext{3}{DSFA, Sezione di Astronomia, Universit\`a di Palermo
Piazza del Parlamento 1, 90134 Palermo, Italy}

\begin{abstract}
An analysis of Lyman series lines arising from hydrogen-like oxygen
and neon ions in the coronae of the active RS~CVn-type binaries II~Peg
and IM~Peg, observed using the {\it Chandra} High Resolution
Transmission Grating Spectrograph, shows significant decrements in the
Ly$\alpha$/Ly$\beta$ ratios as compared with theoretical predictions
and with the same ratios observed in similar active binaries.  We
interpret these decrements in terms of resonance scattering of line
photons out of the line-of-sight; these observations present the first
strong evidence for this effect in active stellar coronae.  The net
line photon loss implies a non-uniform and asymmetric surface
distribution of emitting structures on these stars.  Escape
probability arguments, together with the observed line ratios and
estimates of the emitting plasma density, imply typical line-of-sight
sizes of the coronal structures that dominate the X-ray emission
of $10^{10}$~cm at temperatures of $3\times 10^6$~K and of $10^8$~cm at
$10^7$~K.  These sizes are an order of magnitude larger than predicted
by simple quasi-static coronal loops models, but are still very small 
compared to the several $10^{11}$~cm radii of the underlying stars.
\end{abstract}


\section{Introduction}
\label{s:intro}

The Sun is the only star for which we can presently image coronal
X-ray emission, and in order to understand how coronae on other stars
might be structured we must resort to indirect means.  One potentially
powerful diagnostic of the characteristic size of X-ray emitting
regions is line quenching through resonance scattering.  The escape
probability of a photon emitted by a resonance line in a low density
homogeneous plasma is dependent on the line-of-sight path length
through the plasma region.  Provided the plasma density and the abundance
fraction of the ion in question are known, in the regime where the
plasma is only marginally thick the comparison of strong and weak lines
subject to different scattering losses can yield an estimate of the
typical emitting region path length.

Resonance scattering in the solar corona has been studied
predominantly in the light of the strong ($gf=2.66$) $2p^6\, ^1S_0
\rightarrow 2p^5 3d ^1P_1$ resonance line of Fe~XVII at 15.02~\AA\ as
compared to nearby weaker Fe~XVII lines, though with controversial
results concerning whether optical depth effects were seen or not
(Phillips et al.\ 1996; Phillips et al.\ 1997; Schmelz
et al.\ 1997; Saba et al.\ 1999).
Studies of the same
transition seen in different stars observed at high resolution
($\lambda /\Delta\lambda \sim 1000$) by {\it Chandra} have also
recently been presented by Phillips et al.\ (2001) and Ness et
al.\ (2003).  Both stellar studies failed to find evidence for
significant optical depth. 

There are two problems with using the prominent Fe~XVII soft X-ray
complex for optical depth studies.  Recently, it has been shown by
Doron \& Behar\ (2002) and Gu\ (2003) that the indirect processes of
radiative recombination, dielectronic recombination, and resonance
excitation involving the neighbouring charge states are important for
understanding the relative strengths of Fe~XVII--XX lines.  Secondly,
the coronae of active stars have been found to be Fe-poor by factors
of up to 10 compared with a solar or local cosmic composition (e.g.\ 
reviews by Drake et al.\ 2003, Audard et al.\ 2003), reducing the
sensitivity of Fe lines as optical depth indicators.  Indeed, the
spectral lines likely to exhibit the largest optical depths in the
coronae of active stars are the Lyman $\alpha$ lines of hydrogen-like
O and Ne---elements that are often seen to be enhanced relative to Fe
(e.g.\ Drake et al.\ 2001, Audard et al.\ 2003).

In this {\em Letter}, we present an analysis of
the Ly$\alpha$ to Ly$\beta$ line strength ratios in the active
binaries II~Peg and IM~Peg, and show that the Ly$\alpha$ lines
of Ne and O are significantly quenched.  We use inferred optical
depths to make the first direct estimates of the dimensions of
coronal structures in active stars.

\section{Observations and Analysis}
\label{s:analysis}

Observations and relevant characteristics of the program stars are
summarised in Table~\ref{tab1}.  All spectra were obtained by the {\em
Chandra} High Energy Transmission Grating (HETG) in conjunction with
the Advanced CCD Imaging System spectroscopic detector (ACIS-S), and
were downloaded from the {\em Chandra} Data
Archive\footnote{http://cxc.harvard.edu/cda}.  Spectra from multiple
observations (IM~Peg and AR~Lac) were combined, and positive and
negative orders were summed, keeping HEG and MEG spectra separate.
The same observations of HR~1099, II~Peg and AR~lac are described in
more detail by Drake et al.\ (2001), Huenemoerder et al.\ (2001) and
Huenemoerder et al.\ (2003), respectively.

Resonance scattering of Ly$\alpha$ and Ly$\beta$ photons can be
diagnosed by comparison of the measured Ly$\alpha$/Ly$\beta$ ratio
with respect to the theoretical ratio.  Spectral line fluxes were
measured using the FITLINES utility in the
PINTofALE\footnote{http://hea-www.harvard.edu/PINTofALE/}
IDL\footnote{Interactive Data Language, Research Systems Inc.}
software package (Kashyap \& Drake\ 2000).  In the case of O~VIII, the
Ly$\beta$ transition is blended with an Fe~XVIII line
($2s^22p^5~^2P_{3/2} - 2s^22p^4(^3P)3s~^2P_{3/2}$,
$\lambda=16.004$~\AA).  We estimated its intensity by scaling the
observed intensity of the neighbouring Fe~XVIII 16.071~\AA\ 
($2s^2 2p^5~^2P_{3/2} - 2s^2 2p^4(^3P)3s~^4P_{5/2}$) 
transition, that shares the same upper level, by the ratio of their 
theoretical line strengths (0.76) as predicted by the APED database 
(Smith et al.\ 2001).
Uncertainties involved in this deblending procedure are negligible for
II~Peg (whose Fe~XVIII is weak) and similar to or less than Poisson
errors for the other stars for, e.g., an error of $\sim 20\%$ in the
theoretical Fe XVIII ratio.  He-like resonance lines of Ne~IX and
O~VII were also measured in order to derive a temperature estimate
from the Ly$\alpha$/He-like~$r$ ratio.  Measured line fluxes and
statistical errors are listed in Table~\ref{tab2}.

\begin{figure}[!ht]
\rotatebox{90}{
\epsscale{0.45}
\plotone{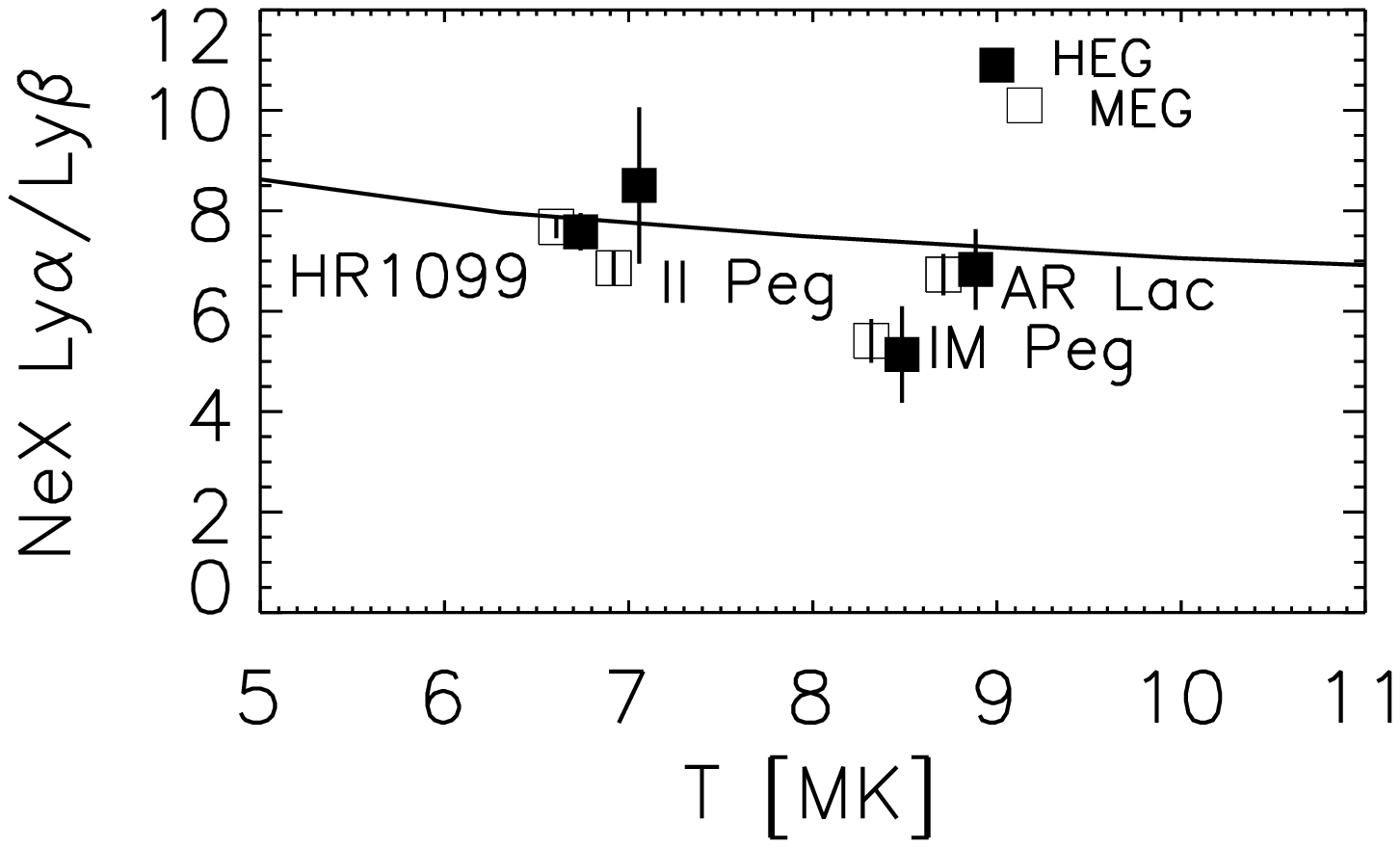}
\plotone{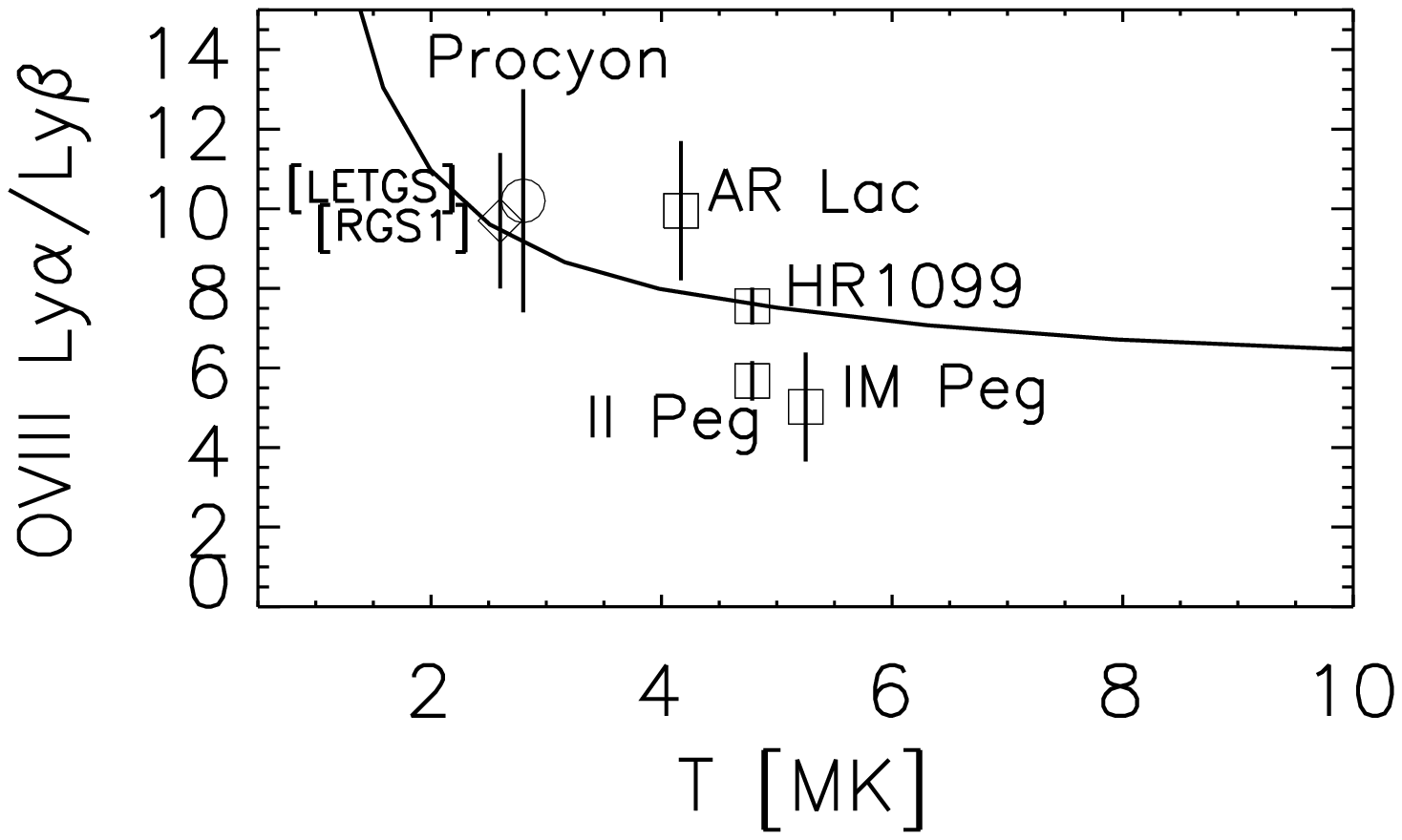}}
\caption{O~VIII ({\em top}) and Ne~X ({\em bottom}) Ly$\alpha/$Ly$\beta$
    ratios for the analyzed sources vs.\ the temperature derived from
    the Ly$\alpha/$He-like~$r$ diagnostics.
    The solid curve is the theoretical ratio from the APED database.
    For the Ne~X we show the results from both the HEG (filled symbols)
    and the MEG (empty symbols) measurements; the HEG ratios are
    shifted by +5\% on the T axis in order to distinguish
    the error bars corresponding to HEG and MEG measurements for the
    same source.
    \label{fig1}}
\end{figure}

Observed and theoretical O and Ne Ly$\alpha$/Ly$\beta$ ratios are
compared in Figure~\ref{fig1}.  The observed ratios are shown at the
temperatures at which observed H-like to
He-like $2p$-$1s$ line intensities matched their theoretical (APED)
values (see Table~\ref{tab3}).  
While it is an approximation to assume that the lines are
formed at a single temperature, this is not critical because the
temperature dependences of these Ly$\alpha$/Ly$\beta$ ratios are not
steep for $T > 2$~MK.  Both O and Ne H-like
lines are formed at temperatures significantly lower than those at
which the theoretical Ly$\alpha$/Ly$\beta$ ratios reach their
asymptotic limits.

The observed Ly$\alpha$/Ly$\beta$ ratios are lower than the
theoretical values for both the Ne~X and the O~VIII lines in the case
of IM~Peg, and for O~VIII in II~Peg.  
Huenemoerder et al.\ (2001) noticed similar
discrepancies in II~Peg but excluded optical depth effects on the grounds that
this deviation was less than $\sim 2\sigma$ from the theoretical
value; it appears, however, that this assessment was based on the high
temperature asymptotic ratio of $\sim 6.25$ (in photon
units).  Our measured ratio for II~Peg is instead more than 3$\sigma$
lower than the expected ratio at the temperature of formation of the
O~VIII lines.  In the case of IM~Peg, the observed O~VIII ratio
lies $\sim 1.8 \sigma$ below the theoretical value, while the Ne~X
ratio in both HEG and MEG, are $>3\sigma$ lower.
For comparison with IM~Peg and II~Peg, we also present similar results
for AR~Lac and HR~1099: these ratios instead do not show significant
departure from the expected optically thin values.

Though uncertainties in the theoretical ratio are not included,
these are not expected to exceed 10\%\ based on good agreement with 
recent laboratory experiments (G.~Brown, private communication; 
Beiersdorfer\ 2003).
If we assume a 10\%\ error in theoretical 
ratios, the Ne Ly$\alpha$/Ly$\beta$ of IM~Peg still departs by 
$3\sigma$ for MEG ($\sim 1.7 \sigma$ for HEG), while the O ratios 
lie $2.5\sigma$ for II~Peg and $1.3\sigma$ for IM~Peg below the 
theoretical curve. 
In Figure~\ref{fig1}, 
we also plot the O~VIII Ly$\alpha$/Ly$\beta$ ratios obtained from
the line fluxes measured by Raassen et al.\ (2002) from 
{\em Chandra}/LETGS and {\em XMM-Newton}/RGS1 spectra of Procyon.
In the Procyon spectrum Fe~XVIII is not detected and these data provide
further validation of the APED O~VIII Ly$\alpha$/Ly$\beta$ ratio.

\begin{figure}[!ht]
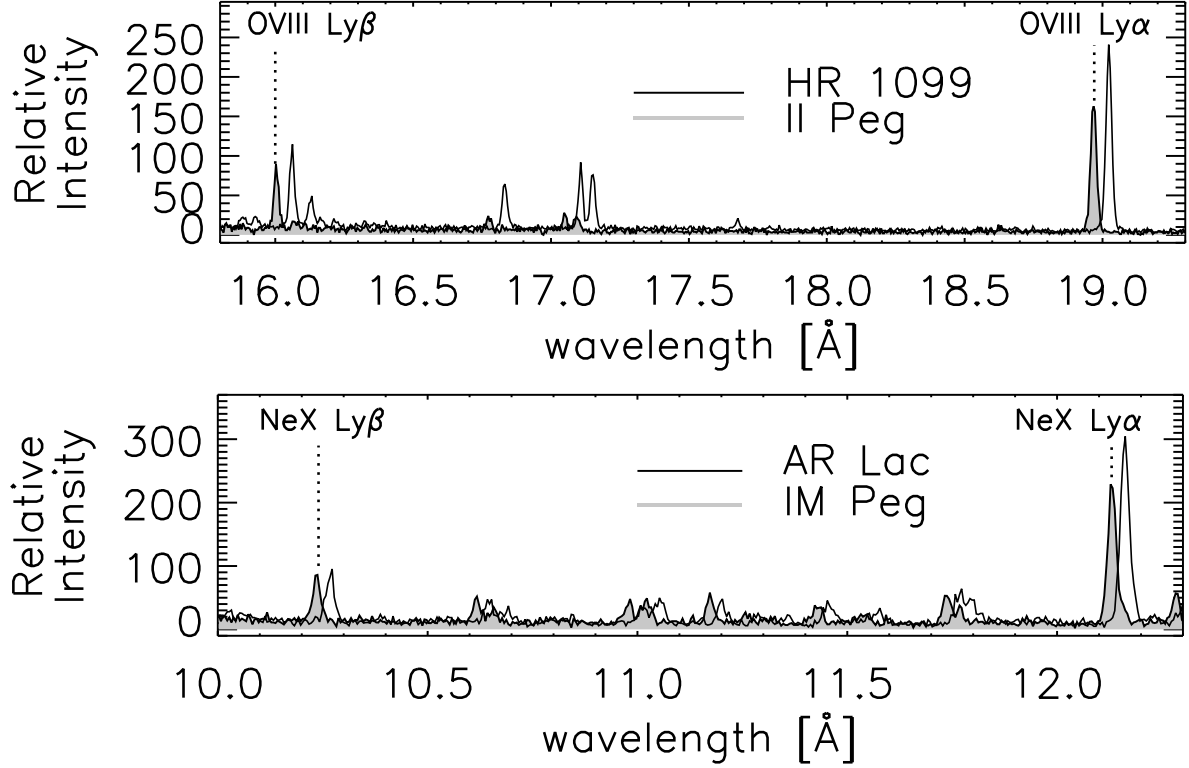

\rotatebox{90}{
\epsscale{0.3}
\plotone{f2b.epsi}
\plotone{f2a.epsi}
}
\caption{Comparison of the spectra for sources showing optical
    depth effects and sources appearing effectively optically thin.
    {\em Top --} O~VIII Ly$\alpha,\beta$ spectral region for
    II~Peg (filled profile) and HR~1099 (shifted in $\lambda$
    by +0.06\AA~for better readability).  Spectra are normalised
    such that they have the same {\em unblended} Ly$\beta$ line 
    strength, i.e.\ after correction for the Fe~XVIII line flux 
    (see text).  The different strengths of the Fe line complex 
    near 17~\AA\ reflects primarily a different Fe/O abundance 
    ratio in the two stars (Drake et al.\ 2001, Huenemoerder et 
    al.\ 2001).
    {\em Bottom --} Ne~X Ly$\alpha,\beta$ spectral region for 
    IM~Peg (filled spectrum) and AR~Lac (shifted in $\lambda$ 
    by +0.03\AA~for better readability). Spectra are again 
    normalized to the observed intensities of the Ly$\beta$ line.
    \label{fig2}}
\end{figure}

These quantitative results are reinforced by a visual comparison of
spectra in Figure~\ref{fig2}, in which the Ly$\alpha$ lines in II~Peg
and IM~Peg are visibly weaker relative to Ly$\beta$ that those of
AR~Lac and HR~1099.  There do not appear to be plausible explanations
for the discrepant ratios other than by the quenching of Ly$\alpha$
relative to Ly$\beta$ through resonance scattering within the emitting
plasma.  Intervening photoelectric absorption could cause similar
effects but H column densities of order a few $10^{21}$~cm$^2$ would
be required---two or three orders of magnitude higher than inferred
for these stars (Mewe et al.\ 1997; Mitrou et al.\ 1997).

\subsection{Path Length Estimate}

Our measured line ratios allow us to derive an effective optical depth
$\tau$, and a typical photon path length within the emitting plasma.
We use the escape probability, $p(\tau)$, derived by Kastner \&
Kastner\ (1990), which for $\tau \lesssim 50$ 
can be approximated by (e.g.\ Kaastra \& Mewe\ 1995)
\begin{equation} 
p(\tau) \sim \frac{1}{1+0.43~\tau}. \label{eq:ptau}
\end{equation}
This is the escape probability for line photons emitted at optical
depths between 0 and $\tau$, averaged over a Gaussian line profile 
due to thermal Doppler broadening, and it assumes that each scattered 
photon is completely lost from the line of sight.  
This probability is significantly larger than the $e^{-\tau}$ 
transmittance of the simple absorption case, since it assumes 
emission over the whole line of sight through the plasma.
The line center optical depth, $\tau$, can be written (e.g., 
Acton\ 1978):
\begin{equation}
 \tau = 1.16 \cdot 10^{-14} \cdot \frac{n_{\rm i}}{n_{\rm el}} 
 	A_{\rm Z} \frac{n_{\rm H}}{n_{\rm e}} \lambda f 
	\sqrt{\frac{M}{T}} n_{\rm e} \ell 
    \label{eq:tau}
\end{equation}
for ion fraction $n_{\rm i}/n_{\rm el}$ (from
Mazzotta et al.\ 1998), element abundance $A_{\rm Z}$, 
oscillator strength $f$, temperature $T$, electron density 
$n_{\rm e}$, atomic weight $M$, where 
$n_{\rm H}/n_{\rm e} \sim 0.85$, and $\ell$ is 
the total path length along the line of sight through the 
emitting plasma. In the above, neglect of non-thermal 
broadening should not be important: non-thermal velocities 
of, e.g., 50~km~s$^{-1}$---similar to the thermal velocities 
for O and Ne at their characteristic temperatures of 
formation---would result in $\ell$ being underestimated by 
a factor of 1.4.

Electron densities were adopted from the survey of Testa et al.\ (2004;
hereafter Paper~I): for $\tau({\rm O~VIII})$ we assumed the
$n_{\rm e}$ derived from the O~VII He-like triplet; the Ne~IX
triplet is strongly affected by blends with Fe lines and so for
$\tau({\rm Ne~X})$ we assumed the $n_{\rm e}$ derived from
the Mg~XI He-like triplet that forms at a similar temperature.  In the
case of IM~Peg, the low signal in the O~VII intercombination and
forbidden lines precludes a firm density estimate; we therefore
assumed a value of $2\times 10^{10}$~cm$^{-3}$, which is typical of
values obtained for all the measurable spectra of RS~CVns in Paper~I.
The assumed $n_{\rm e}$ are listed in Table~\ref{tab3}, while the 
adopted O and Ne abundances are listed in Table~\ref{tab4}.

In order to derive an estimate of the path length, $\ell$, we treated
the fine structure components of Ly$\alpha$ (1:~$^2P_{3/2} \rightarrow
^2S_{1/2}$ at $\lambda=18.9671$~\AA\ and 2:~$^2P_{1/2} \rightarrow
^2S_{1/2}$ at $\lambda=18.9726$~\AA) separately, since their splitting
($\Delta \lambda/ \lambda \sim 0.0003$) is larger than the thermal
width ($\sim \sqrt{k_BT/M}/c \sim 0.00013$).  The observed and
theoretical intensities, $I_{\rm obs}$ and $I_{\rm th}$, 
are then related as follows:

\begin{equation}
I_{Ly\alpha~{\rm obs}} = I_{1~{\rm obs}}+ I_{2~{\rm obs}} =
    \frac{I_{1~{\rm th}}}{1+0.43 \tau_1}
    + \frac{I_{2~{\rm th}}}{1+0.43 \tau_2} .
\end{equation}
Since for hydrogenic ions $f_2=f_1/2=0.2776/2$ (e.g.\ Morton\ 
2003), $I_{2~{\rm th}}=I_{1~{\rm th}}/2$,
and $\tau_i \sim C(\ell) \cdot f_i$
(see Eq.~\ref{eq:tau}), we obtain:

\begin{equation} 
\frac{I_{Ly\alpha~{\rm obs}}}{I_{Ly\alpha~{\rm th}}} = \frac{1}{3}
     \left[\frac{2}{1+0.43 C(\ell) f_1}
    + \frac{1}{1+0.43 C(\ell) f_1/2}\right] .
\end{equation}
For Ly$\beta$ the splitting of the components is smaller than the 
thermal width and the escape probability for Ly$\beta$ is given 
directly by Eq.\ref{eq:ptau}.
The combined equation for $C(\ell)$ was then solved to obtain the 
path length.

The path length estimates, $\ell_{\tau}$, derived from the
measurements are listed in Table~\ref{tab4}.
For comparison, we also list the stellar radii and loop lengths 
expected for a standard hydrostatic loop model (e.g.\ Rosner et
al.\ 1978, RTV hereafter), $L_{\rm RTV}$, corresponding to the 
observed temperatures and densities.

\section{Discussion and Conclusions}
\label{s:discuss}

The discrepant O~VIII and Ne~X Ly$\alpha$/Ly$\beta$ ratios found here
for II~Peg and IM~Peg represent the first clear evidence of resonant
scattering in coronal X-ray emission lines. 
The photon path lengths inferred from the observed ratios
(Table~\ref{tab4}) are: 
(1) about two orders of magnitude different from each other, 
reflecting the differences in the plasma densities found in Paper~I 
for the characteristic temperatures of formation of the O~VIII and 
Ne~X lines;
(2) very small with respect to the stellar radius;
(3) an order of magnitude larger than the loop lengths derived 
from RTV model scaling laws.   In the ratio
$\ell_{\tau}/L_{\rm RTV}$ the $n_e$ terms cancel, so that this 
conclusion is completely independent of plasma density measurements.

We note that the path lengths obtained from this type of analysis
should be treated as lower limits because addition of Ly$\alpha$ 
photons by any optically thin emission or any scattering
into the line-of-sight are not taken into account.  In the case of a
uniform spherically-symmetric arrangement of emitting structures over
the stellar surface, scattering into and out of the line-of-sight
would be expected to compensate one another.  The detection of
resonance scattering implicitly suggests a non-uniform coronal
distribution.

That we find different optical depths in emitting regions on what are
ostensibly similar active RS~CVn-type binaries is puzzling.  One
possibility is that this is transient behaviour that might occur on
all similarly active stars for a preferential arrangement of one or
more coronal structures.  We also note that, whereas both AR~Lac and
HR~1099 comprise two late-type stars of more similar type, II~Peg and
IM~Peg have unseen companions of unknown spectral type, though how
this arrangement might benefit larger coronal photon path lengths 
is not obvious.

\begin{acknowledgements}
We thank the referee, Jeffrey L. Linsky, for helpful comments
that enabled us to improve the manuscript.
PT was partially supported by {\it Chandra} grants GO1-20006X
and GO1-2012X under the SAO Predoctoral Fellowship program.  
JJD was supported by NASA contract NAS8-39073 to the
{\em Chandra X-ray Center}; EED was supported by NASA grant 
NAG5-10872.  GP and PT were partially supported by MIUR and 
by ASI.
\end{acknowledgements}

\clearpage

\input{tab1}

\clearpage

\input{tab2}

\clearpage

\input{tab3}

\clearpage

\input{tab4}

\clearpage

\clearpage

\end{document}

%% file: tab1.tex
\begin{deluxetable}{lcccccccccc}
\rotate
\tabletypesize{\footnotesize}
\tablecaption{Properties of Observed Stars and 
	related HETG observations. \label{tab1}}
\tablewidth{0pt}
\tablehead{
 \colhead{Source} & \colhead{Spectr.Type} & 
 \colhead{d \tablenotemark{(1)}} & 
 \colhead{R$_{\star}$/R$_{\odot}$} & 
 \colhead{P$_{\rm orb}$ \tablenotemark{(2)}} &
 \colhead{P$_{\rm rot}$ \tablenotemark{(2)}} &
 \colhead{L$_{\rm X}$ [HEG]\tablenotemark{a}}  &
 \colhead{L$_{\rm X}$ [MEG]\tablenotemark{a}}  &
 \colhead{F$_{\rm X}$ \tablenotemark{b}} &
 \colhead{Obs ID} & \colhead{$t_{\rm exp}$} \\
  & & [pc] & & [ksec] & [ksec] & [erg/sec] &
 [erg/sec] & [$10^5$ erg/cm$^2$/sec] &  & [ksec] 
}
\startdata
 II~Peg	 & K2V/..  & 42  & 3.4/.. \tablenotemark{(3)} & 580 
 	 & 580 & 1.56$\cdot 10^{31}$ &  1.76$\cdot 10^{31}$ 
	 & 250 & 1451 & 42.7 \\  
 IM~Peg	 & K2III-II/.. & 96.8  & 13/.. \tablenotemark{(4)} 
 	 & 2100 & 2100 & 2.75$\cdot 10^{31}$ 
	 & 2.79$\cdot 10^{31}$ & 27.1 & 2527 & 24.6 \\ 
 IM~Peg  & & & & & & 2.17$\cdot 10^{31}$ 
 	 & 2.30$\cdot 10^{31}$ & 22.3 & 2528 & 24.8 \\ 
 IM~Peg  & & & & & & 1.86$\cdot 10^{31}$ 
 	 & 1.97$\cdot 10^{31}$ & 19.2 & 2529 & 24.8 \\  
 AR~Lac	 & G2IV/K0IV & 42  & 1.8/3.1 \tablenotemark{(2)} 
 	 & 170 & 170 & 5.21$\cdot 10^{30}$ 
	 & 5.60$\cdot 10^{30}$ & 284 & 6 & 32.1 \\   
 AR~Lac	 & & & & & & 5.61$\cdot 10^{30}$ 
 	 & 6.30$\cdot 10^{30}$ & 319 & 9 & 32.2 \\   
 HR~1099 & G5IV/K1IV & 29.0 & 1.3/3.9 \tablenotemark{(2)} 
 	 & 250 & 250 & 7.85$\cdot 10^{30}$ 
	 & 1.05$\cdot 10^{31}$ & 1020 & 62538 & 94.7 
\enddata 
\tablenotetext{a}{ HEG range: 1.5-15~\AA ; MEG range: 2-24~\AA.}
\tablenotetext{b}{ X-ray surface flux from L$_{\rm X}$
	obtained from MEG spectra}
\tablecomments{References:  $^{(1)}$ SIMBAD database;
	$^{(2)}$ Strassmeier et al.\ 1993; 
	$^{(3)}$ Berdyugina et al.\ 1998; 
 	$^{(4)}$ Berdyugina et al.\ 1999}
\end{deluxetable}

%% file: tab2.tex
\begin{deluxetable}{lcrrrrrrcccc}
\rotate
\tablecolumns{12} 
\tabletypesize{\footnotesize}
\tablecaption{Line flux measurements and 
	Ly$\alpha$/Ly$\beta$ ratios, with 1$\sigma$ errors.
	 \label{tab2}}
\tablewidth{0pt}
\tablehead{
 \colhead{Source} & \colhead{grating} &
 \multicolumn{7}{c}{flux  
 ($10^{-6}$~photons~cm$^{-2}$~sec$^{-1}$)} & 
 \colhead{} & \multicolumn{2}{c}{Photon Ratio}\\
 \cline{3-9} \\[-0.15cm]
 \colhead{} & \colhead{} &
 \multicolumn{2}{c}{Ne~X} & \colhead{Ne~IX} &
 \multicolumn{2}{c}{O~VIII} & \colhead{O~VII} &
 \colhead{Fe~XVIII} & \colhead{} & 
 \multicolumn{2}{c}{Ly$\alpha$/Ly$\beta$} \\
 \cline{3-4} \cline{6-7} \cline{11-12} \\[-0.15cm]
 \colhead{} & \colhead{} & \colhead{Ly$\alpha$} & 
 \colhead{Ly$\beta$} & \colhead{$r$} & 
 \colhead{Ly$\alpha$} &  \colhead{Ly$\beta$} &
 \colhead{$r$} & \colhead{} & \colhead{} & 
 \colhead{Ne~X} & \colhead{O~VIII} \\
 \colhead{} & \colhead{} & \colhead{12.132\AA} & 
 \colhead{10.239\AA} & \colhead{13.447\AA} & 
 \colhead{18.967\AA} & \colhead{16.006\AA} &
 \colhead{21.602\AA} & \colhead{16.071\AA} &
 \colhead{} & \colhead{} & \colhead{}
}
\startdata 
 II~Peg  & HEG & 1364$\pm$70 & 160$\pm$28 & & & & & & 
 	 & $8.5 \pm 1.6$ & \\ 
  	 & MEG & 1217$\pm$28 & 178$\pm$8 & 365$\pm$21 
	 & 2050$\pm$70 & 419$\pm$26 & 250$\pm$50 & 77$\pm$16 
	 & & $6.8 \pm 0.3$ & $5.6 \pm 0.5$ \\[0.1cm]
 IM~Peg  & HEG & 410$\pm$30 & 80$\pm$20 &  & & & & & 
 	 & $5.1 \pm 0.9$ & \\ 
  	 & MEG & 385$\pm$13 & 71$\pm$2 & 77$\pm$11 
	 & 530$\pm$40 & 145$\pm$22 & 52$\pm$25 & 55$\pm$13 
	 & & $5.4 \pm 0.4$ & $5.0 \pm 1.2$ \\[0.1cm]
 AR~Lac  & HEG & 727$\pm$30 & 106$\pm$11 & & & & & &
 	 & $6.8 \pm 0.8$ &  \\ 
  	 & MEG & 632$\pm$16 & 94$\pm$5 & 107$\pm$14 
	 & 820$\pm$40 & 175$\pm$15 & 156$\pm$44 & 121$\pm$14
	 & & $6.7 \pm 0.4$ & $9.9 \pm 1.9$   \\[0.1cm]
 HR~1099 & HEG & 2180$\pm$40 & 288$\pm$13 & & & & & &
 	 & $7.6 \pm 0.4$ &  \\ 
  	 & MEG & 1705$\pm$21 & 222$\pm$6 & 571$\pm$16 
	 & 2860$\pm$60 & 549$\pm$19 & 400$\pm$40 
	 & 225$\pm$14 & & $7.7 \pm 0.2$ & $7.6 \pm 0.5$
\enddata
\end{deluxetable}

%% file: tab3.tex
\begin{deluxetable}{lrccccc}
\tablecolumns{7} 
\tabletypesize{\footnotesize}
\tablecaption{Plasma densities and Temperatures. 
	\label{tab3}}
\tablewidth{0pt}
\tablehead{
 \colhead{} & \colhead{} & \colhead{} & \colhead{II~Peg} & 
 \colhead{IM~Peg} & \colhead{AR~Lac} & \colhead{HR~1099} 
}
\startdata 
 $n_{\mathrm{e}}$ \tablenotemark{a} & Mg~XI & 
  ($10^{12}$cm$^{-3}$) & $5.6^{+0.6}_{-2.5}$ & 
  $3.2^{+2.5}_{-1.4}$  & $<1.8$ & $1.8^{+0.6}_{-0.5}$ \\[0.1cm]
 $n_{\mathrm{e}}$ \tablenotemark{a} & O~VII & 
  ($10^{10}$cm$^{-3}$) & $3.2^{+6.8}_{-1.4}$ & 
  - & - & $1.0^{+2.2}_{-0.8}$ \\[0.1cm] \hline
 $T$ \tablenotemark{b} & Ne & (MK) & 6.7-7.3 & 7.8-8.8 
 		& 8.1-9.6 & 6.0-6.4 \\
 $T$ \tablenotemark{b} & O  & (MK) & 4.7-5.5 & 4.5-7 
 		& 4-5 & 4.7-5 
\enddata
\tablenotetext{a}{ From He-like triplet diagnostics (Paper~I).}
\tablenotetext{b}{ From Ly$\alpha$/r ratio diagnostics.}
\end{deluxetable}

%% file: tab4.tex
\begin{deluxetable}{llcccl}
\tabletypesize{\footnotesize}
\tablecaption{Path length derived from measured 
	Ly$\alpha$/Ly$\beta$.
	\label{tab4}}
\tablewidth{0pt}
\tablehead{
 \colhead{Source} & \colhead{Ion} & 
 \colhead{Element\tablenotemark{a}} & \colhead{$\ell_{\tau}$} 
 & \colhead{$L_{\rm RTV}$ \tablenotemark{b}}
 & \colhead{$\ell_{\tau}$/R$_{\star}$ \tablenotemark{c}} \\
 \colhead{} & \colhead{} & \colhead{Abundance} & 
 \colhead{[cm]}  &  \colhead{[cm]} & \colhead{}
}
\startdata
 II~Peg	 & O~VIII & 8.97\tablenotemark{d} &  $9.5 \cdot 10^{9}$  
 	 & $1 \cdot 10^{9}$  & 0.04 \\[0.15cm]
 IM~Peg  & O~VIII &  9.37\tablenotemark{e} &  $1.7 \cdot 10^{10}$   
 	 & $2.2 \cdot 10^{9}$  & 0.019 \\  
	 & Ne~X [HEG] & 8.86\tablenotemark{e} & $1.6 \cdot 10^{8}$  
	 & $2.8 \cdot 10^{7}$ & 0.0002 \\  
	 & Ne~X [MEG] & \nodata & $2.2 \cdot 10^{8}$  
	 & $2.8 \cdot 10^{7}$ & 0.00018 
 \enddata 
\tablenotetext{a}{ Expressed on the usual spectroscopic logarithmic 
	scale where X/H=log(n(X)/n(H))+12, and n(X) is the number 
	density of the element X.}
\tablenotetext{b}{ {Loop} length from RTV scaling laws:
   	$L_{\rm RTV} \sim T^2/[(1.4 \cdot 10^3)^3 \cdot 2n_{\rm e}k_B]$}
\tablenotetext{c}{ Path length as fraction of the stellar radius}
\tablenotetext{d}{ From Huenemoerder et al.\ (2001)}
\tablenotetext{e}{ Scaled from Huenemoerder et al.\ (2001) II~Peg
	values by the ratio of II~Peg to IM~Peg photospheric
	metallicities, [Fe/H]$_{\rm II}=-0.4$ and [Fe/H]$_{\rm IM}=0.0$,
	derived by Berdyugina et al.\ (1998,1999, respectively)}
\end{deluxetable}